\documentclass[times,amssymb,usenatbib,useAMS,twocolumns]{mnras}
\usepackage{times} 
\usepackage[utf8]{inputenc}
\usepackage{amsmath}
\usepackage{amsfonts}
\usepackage{graphicx}
\usepackage{ragged2e}
\justifying

\title[Deuteronated PAHs]{Theoretical study of deuteronated PAHs as carriers for IR emission features in the ISM}
\author[Buragohain et al.]
{
\parbox{\textwidth}{
Mridusmita Buragohain$^{1}$, 
Amit Pathak$^{1}$, 
Peter Sarre$^2$, 
Takashi Onaka$^{3}$ and 
Itsuki Sakon$^{3}$}
\vspace*{10pt}\\
$^{1}$Department of Physics, Tezpur University, Tezpur 784\,028, India (ms.mridusmita@gmail.com, amit@tezu.ernet.in)\\
$^{2}$School of Chemistry, The University of Nottingham, University Park, Nottingham, NG7~2RD, United Kingdom\\
$^{3}$Department of Astronomy, Graduate School of Science, The University of Tokyo, Tokyo 113-0033, Japan}

\begin{document}
\date{}
\maketitle

\begin{abstract}
This work proposes deuteronated PAH (DPAH$^+$) molecules as a potential carrier
of the 4.4 and 4.65$~\mu \rm m$ mid infrared emission bands that have been
observationally detected towards the Orion and M17 regions. Density Functional
Theory calculations have been carried out on DPAH$^+$ molecules to see the
variations in the spectral behaviour from that of a pure PAH. DPAH$^+$ molecules
show features that arise due to the stretching of the aliphatic C-D bond.
Deuterated PAHs have been previously reported as carriers for such features.
However, preferred conditions of ionization of PAHs in the interstellar medium
(ISM) indicates the possibility of the formation of DPAH$^+$ molecules.
Comparison of band positions of DPAH$^+$s shows reasonable agreement with the
observations. We report the effect of size of the DPAH$^+$ molecules on band
positions and intensities. This study also reports a D/H ratio ([D/H]$\rm_{sc}$;
the ratio of $\rm C-D$ stretch and $\rm C-H$ stretch bands per [D/H]$\rm_{num}$) that is
decreasing with the increasing size of DPAH$^+$s. It is noted that large
DPAH$^+$ molecules (no. of C atoms $\sim$~50) match the D/H ratio that has been estimated from observations.
This ratio offers prospects to study the deuterium abundance and depletion in
the ISM.
\end{abstract}



\section{Introduction}

In recent years, the detection of unidentified infrared (UIR) emission bands at 3.3, 6.2, 7.7, 8.6, 11.2, 12.7 and $16.4~\mu \rm m$ 
towards many astronomical objects in the interstellar medium (ISM) has opened up new prospects in observational, laboratory and 
theoretical molecular astrophysics. These bands were first detected by \citet[]{Gillett73} as broad emission features together 
with line as well as continuum radiation towards three planetary nebulae NGC 7027, BD+30$\rm^o$3639 and NGC 6572. 
\textit{ISO (SWS)} and \textit{Spitzer} observations have revealed that these features are ubiquitous in the ISM 
\citep[]{Cox99, Smith07}. With the progress in observational astronomy, these features have further been observed 
towards a variety of astronomical objects including H~\textsc{ii} regions, reflection nebulae, planetary nebulae, 
photodissociation regions, AGB objects, active star forming regions, young stellar objects, diffuse medium, etc. 
\citep[]{Onaka96, Mattila96, Verstraete96, Moutou99, Hony01, Verstraete01, Peeters02, Abergel02, Acke04, Sakon04}. 
These features are also observed in external galaxies of varying metallicity \citep[]{Genzel98, Mattila99, 
Helou2000, Lu03, Regan04, Brandl06, Armus07, Onaka08, Mori12}. Apart from strong UIR band emission, there 
are weak broad features distributed in the emission plateaux in the $\sim3-20~\mu \rm m$ range 
\citep[]{Tielens08}. Depending on the physical conditions of the observed environment, source to source 
variation of the observed IR features is noticed in terms of peak position, width and intensity, however, 
an overall correlation among the bands is maintained \citep[]{Hony01, Peeters02, Sakon07, Tielens08, Mori12}.

It was first proposed by \citet[]{Leger84} and by \citet[]{Allamandola85} that these features arise due to 
excitation of polycyclic aromatic hydrocarbon (PAH) molecules. When a PAH molecule absorbs an UV photon, it
is either ionized or it gets excited to a higher electronic state. These excited PAHs tend to relax 
through radiation-less processes, including dissociation, internal conversion and intersystem crossing. As
a consequence, the molecule comes to its ground electronic state while still in a vibrationally 
excited state. The molecule emits through IR emission which reverts it back to its ground state. The excess
energy is released off through different modes of vibration, giving rise to IR emission features 
\citep[]{Allamandola89, Puget89}. Though the widespread presence of PAHs has been established by the
observation of UIR bands, the assignment of the exact PAH form responsible for these bands still presents 
some challenges. Several weaker bands have been detected by Short Wavelength Spectrometer (\textit{SWS}) 
on board the \textit{ISO} satellite \citep[]{Verstraete96, Peeters04}. In order to assign carriers for these bands, 
it is essential to compare the observational data with experimental as well as theoretical spectral data.

PAHs constitute a significant fraction of the material in the ISM bearing about 5-10\% of the elemental 
carbon \citep[]{Tielens08} and contribute in various interstellar processes like heating of the ISM through
the photoelectric effect and influence on the charge balance inside molecular clouds \citep[]{d'Hendecourt87,Lepp88, 
Verstraete90, Bakes94, Peeters04}. It is important to study the formation of PAHs in astrophysical environments 
in order to understand the chemical processes occurring in the ISM. PAHs have also been proposed to be the 
carriers of the Diffuse Interstellar Bands (DIBs) observed towards Galactic and extragalactic sources
\citep[]{Crawford85, Leger85, Salama96, Salama11, Cox06}. Interstellar PAHs encompass a vast range of 
families and a single category of PAH alone cannot explain the complete set of UIR bands. Several 
substituted PAHs have been studied theoretically and experimentally in order to seek to match the 
UIR bands. These include nitrogen substituted PAHs \citep[]{Cook96, Bauschlicher98a, Mattioda03, 
Mattioda05, Hudgins04, Hudgins05, Galue10}, oxygen substituted PAHs \citep[]{Cook96, Bauschlicher98b, 
Hudgins05}, silicon substituted PAHs \citep[]{Hudgins05}, methyl substituted PAHs \citep[]{Bauschlicher98c}, 
and PAHs with Fe and Mg \citep[]{Serra92, Klotz95, Hudgins05, Simon11}.

Among the substituted PAHs, deuterated PAHs (PADs) have been studied extensively both experimentally and
theoretically in relation to the UIR bands \citep[]{Bausch97a, Hudgins04a}. PADs show distinct features 
at 4.4 and 4.65$~\mu \rm m$ that have been observed towards Orion Nebula and M17 \citep[]{Peeters04}. These observations
 have not been confirmed by \textit{AKARI} observations \citep[]{Onaka14}. 
These emission bands are characteristics of $\rm{C-D}$ stretching modes in PADs \citep[]{Peeters04}. 
Deuterium (D) detection in the ISM established by the higher Lyman lines seen in \textit {FUSE} spectra \citep[]{Hoopes03} 
further justifies the existence of PADs in the ISM. Interstellar PAHs may become deuterium enriched by exchange 
of D from D$_2$O ice when exposed to UV radiation \citep[]{Sandford2000}. The 4.4 and 4.65$~\mu \rm m$ bands 
provide a good prospect for observational searches for PADs as they do not overlap with any other PAH features
and are a pure $\rm{C-D}$ contribution \citep[]{Hudgins04a}. Interstellar PADs are proposed to be a major 
reservoir for D which may explain the present D/H ratio in the ISM \citep[]{Draine06}. \citet[]{Linsky06} 
discussed a variation in D/H ratios along various lines of sight in the Milky Way and beyond. According to 
the `Deuterium depletion model', \citet[]{Draine06} proposed a D/H ratio of $\sim$~0.3 in PAHs. 
A similar ratio is obtained by \citet[]{Peeters04}. However, \textit{AKARI} data did not confirm this and 
does not support the presence of large amount of PADs in the ISM \citep[]{Onaka14}.

In this work we consider another class of deuterium enriched PAHs, deuteronated PAHs 
(DPAH$^+$)~\footnote{deuteronated PAHs are PAHs to which a deuteron is added - the equivalent of protonated
PAHs for a proton.}. DPAH$^+$ molecules might be crucial in an astrophysical context because their closed-shell
electronic structure  makes them stable enough to survive the extreme interstellar environment. Here, we report 
theoretical vibrational calculations of DPAH$^+$ as probable carriers of some UIR bands and make comparison
with observed UIR features.

\section{Probable DPAH$^+$ formation mechanism in the ISM}

Deuterium is considered to be one of the lightest elements formed after the Big Bang and in the chemical evolution 
of the Universe, it is converted to heavier elements by nuclear fusion in stellar interiors. The present lower 
value of D/H ($\sim$~7ppm to $\sim$~22 ppm); \citep[]{Jenkins99, Sonneborn2000, Wood04} compared to the primordial 
one ($\sim$26 ppm) \citep[]{Epstein76, Mazzitelli80, Moos02, Steigman03, Wood04} is explained by this. However, 
\citet[]{Draine06} argued that it is highly possible that some primordial deuterium is depleted in interstellar 
dust. \citet[]{Draine06} further suggested that of all the interstellar dust grains incorporating deuterium, some
may be in PAHs that may result in the formation of PADs. Deuterated PAHs (PADs or D$\rm{_n}$-PAHs) may be formed in 
the ISM by the following chemical processes \\
i) gas-phase ion-molecule reactions in low temperature environments \citep[]{Tielens97}, 
ii) gas-grain reactions \citep[]{Tielens83, Tielens92, Tielens97}, iii) photodissociation of carbonaceous 
dust grains \citep[]{Allamandola87, Allamandola89} and iv) exchange of deuterium in D$_2$O ice with one of 
the peripheral hydrogen atoms of an interstellar PAH when exposed to UV radiation \citep[]{Sandford00}.

Formation of DPAH$^+$ molecules may result from a number of interstellar processes. These include:

(i) Addition of D to PAH radical cations:
\begin{equation*}
 \rm PAH^{+} + D \rightarrow DPAH^{+}
\end{equation*}

\citet[]{LePage97}, \citet[]{Snow98}, and \citet[]{LePage01} discussed the reaction of PAH cations with atomic and molecular 
hydrogen to form protonated PAHs (HPAH$^+$) that might emerge as a potential carriers of DIBs \citep[]{Pathak08}. Larger 
PAH cations tend to associate efficiently with atomic H to form HPAH$^+$ in interstellar environments \citep[]{Herbst99}.
In a similar way, deuterium atoms may also react with PAH cations to form deuteronated PAHs.

(ii) Addition of D$^+$, produced by direct ionization and charge-transfer reaction, to a neutral PAH to form DPAH$^+$:
\begin{equation*}
\rm H^{+} + D \rightleftharpoons D^{+} + H~(\rm charge-transfer~reaction)
\end{equation*}
\begin{equation*}
\rm D^+ + PAH \rightarrow DPAH^+ + h\nu
\end{equation*}

(iii) Low temperature ion-molecule reaction followed by deuterium fractionation: At temperatures below 50\,K, deuterium 
fractionation is significant \citep[]{Millar89}. Interstellar deuterium mostly exists in the form of HD. Fractionation of
HD by exchange reaction with H$_3^+$ occurs efficiently in low temperature dense interstellar clouds, to form deuterated 
molecular ions H$_2$D$^+$ \citep[]{Millar89}. H$_2$D$^+$ has a low deuteron affinity \citep[]{Roberts02} and can 
transfer D$^+$ to PAH to form DPAH$^+$.
\begin{equation*}
\rm H_{3}^{+} + HD \rightleftharpoons H_{2}D^{+} + H_{2}
\end{equation*}
\begin{equation*}
\rm H_{2}D^{+} + PAH \rightarrow DPAH^{+} + H_{2}
\end{equation*}

(iv) D may merely replace the hydrogen at the protonation site in HPAH$^+$ without altering the network to form DPAH$^+$
\begin{equation*}
\rm D + HPAH^{+} \rightarrow DPAH^{+} + H
\end{equation*}

Due to the higher mass of deuterium in DPAH$^+$, such interstellar species are expected to give spectral modes (associated with D) 
towards longer wavelengths compared to their neutral or protonated counterparts.

\section{Computational approach}

Theoretical quantum chemical calculations help in narrowing down candidates for much more expensive laboratory experiments. 
Considering the high cost, time consumption and other constraints faced in laboratory, theoretical computational study 
can propose selected PAHs for which laboratory spectroscopy can most usefully be performed. Density Functional Theory 
(DFT) has been used rigorously to calculate the harmonic frequencies and intensities of vibrational modes of PAHs in 
various forms including size, composition and charge states \citep[]{Langhoff96, Bausch97a, Bausch97b, Langhoff98, 
Hudgins01, Hudgins04a, Pathak05, Pathak06, Pathak07, Pathak08, Candian14}. In the present work, DFT in combination 
with a B3LYP functional and a 6-311G** basis set has been used to optimize the molecular structures of PAHs. The 
optimized geometry is used to obtain the vibrational frequencies of various modes at the same level of theory. 
Theoretical calculations tend to overestimate the frequency compared to experiments \citep[]{Langhoff96}. 
The use of a larger basis set, \emph{e.g.} 6-311G**, generally reduces the overestimation compared to smaller
basis sets \citep[]{Langhoff96}. Use of a larger basis set compared to a smaller one also shows good agreement
with experiment. However, the use of a large basis set does not support use of a single scaling factor for all
of the vibrational modes \citep[]{Langhoff96}. In order to evaluate the mode-dependent scaling factors, 
calibration calculations were made for selected PAHs, both neutral and ionized. On comparing the theoretical 
frequencies with matrix isolated spectroscopic experimental data \citep[]{Hudgins95a, Hudgins95b, Hudgins98a, 
Hudgins98b}, three different scaling factors have been determined. The scaling factors obtained are 0.974 for 
the $\rm{C-H}$ out-of-plane (oop) mode, 0.972 for the $\rm{C-H}$ in-plane and $\rm{C-C}$ stretching modes and 
0.965 for the $\rm{C-H}$ stretching mode. Gaussian line shapes of 30~$\rm cm^{-1}$ FWHM are used to plot the 
computationally obtained spectra. Our sample includes deuteronated pyrene (DC$_{16}$H$_{10}^+$), deuteronated 
perylene (DC$_{20}$H$_{12}^+$) and deuteronated coronene (DC$_{24}$H$_{12}^+$). Isomers of DC$_{16}$H$_{10}^+$ 
and DC$_{20}$H$_{12}^+$ have also been included. The data presented here were produced using GAMESS quantum 
chemistry suite of programs \citep[]{Schmidt93}.

\section{Results and Discussion}


\subsection*{Deuteronated pyrene}

\begin{figure*}
\centering
\includegraphics[width=10cm,height=12cm]{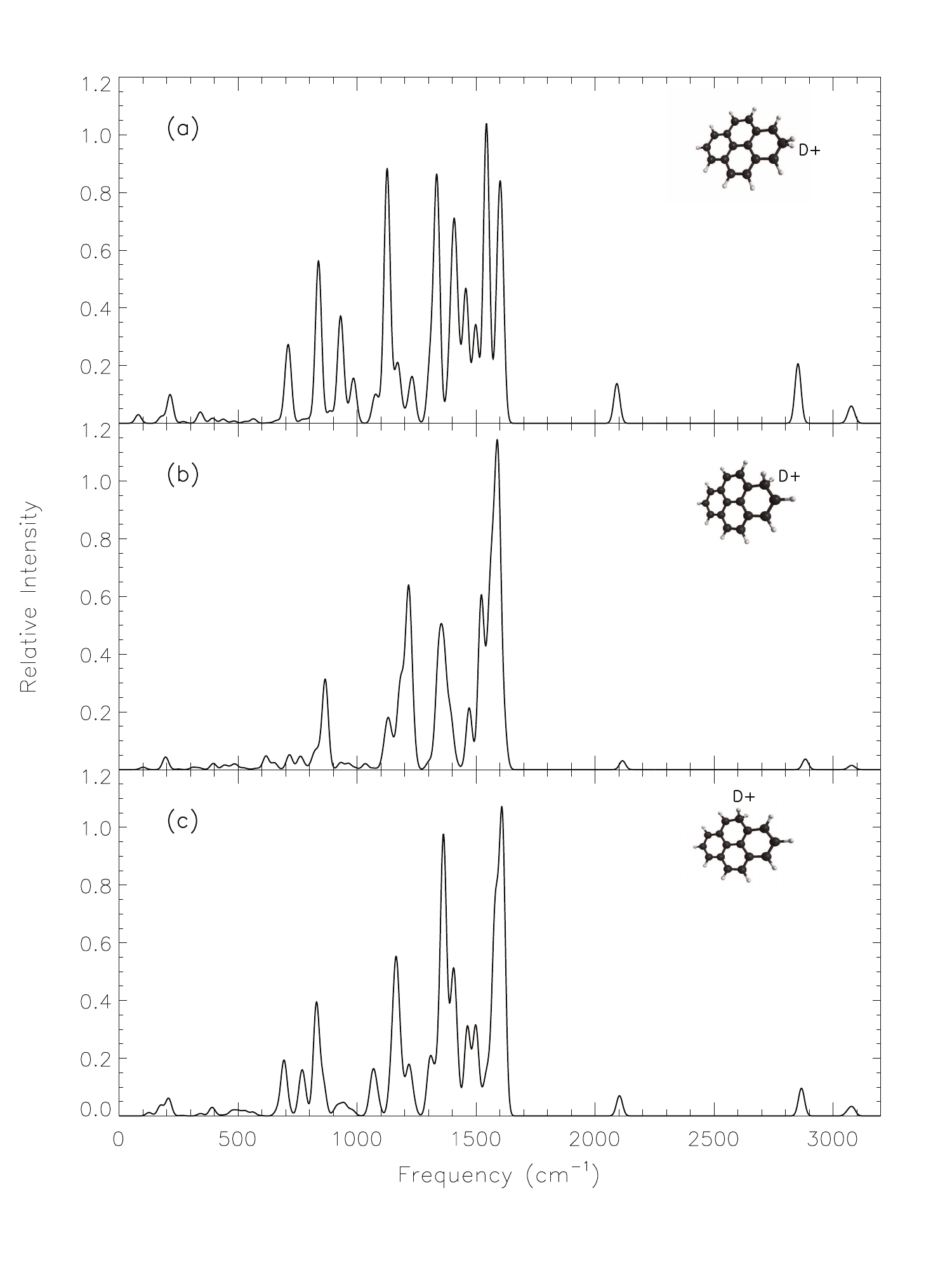}
\caption{ Theoretical spectra of (a) DC$_{16}$H$_{10}^+$, (b) Isomer 1 of DC$_{16}$H$_{10}^+$, (c) Isomer 2 of DC$_{16}$H$_{10}^+$}
\label{fig1}
\end{figure*}

Fig.~\ref{fig1} shows the theoretical spectra of deuteronated pyrene (DC$_{16}$H$_{10}^+$) and its isomers. 
Each isomer shows a unique spectrum. Deuteronated pyrene with C$\rm_{2v}$ symmetry (Fig.~\ref{fig1}a) shows 
pronounced transitions particularly in the 825-1600~$\rm cm^{-1}$ (12.12-6.25~$\mu \rm m$) range compared to 
its isomers with C$\rm_s$ symmetry (Fig.~\ref{fig1}b and Fig.~\ref{fig1}c). Most of the bands are characteristic 
of $\rm{C-H}$ out-of-plane (oop), $\rm{C-H}$ in-plane, $\rm{C-C}$ stretching and $\rm{C-H}$ stretching modes. 
However, some new features arise due to the contribution of D. These include $\rm{C-D}$ in-plane, $\rm{H-C-D}$ 
oop and $\rm{C-D}$ stretching particularly. The $\rm{C-D}$ in-plane modes are distributed in 
the $\sim~700-870~\rm cm^{-1}$ (14-11~$\mu \rm m$) range with varying relative intensities between 
0.05 to 0.4. This is the same region where $\rm{C-H}$ oop occurs in unsubstituted PAHs. The $\rm{H-C-D}$ oop 
modes are relatively strong and present in the narrow range of 1125-1185~$\rm cm^{-1}$ (8.4-8.9~$\mu \rm m$). 
The $\rm{C-D}$ in-plane and $\rm{H-C-D}$ oop modes are found to be blended with other fundamental modes. A 
significant aliphatic $\rm{C-D}$ stretching mode appears at 2092~$\rm cm^{-1}$ (4.78~$\mu \rm m$) in Fig.~\ref{fig1}a 
and faintly in the other two isomers at 2116~$\rm cm^{-1}$ (4.73~$\mu \rm m$) and 2103~$\rm cm^{-1}$ 
(4.76~$\mu \rm m$) (Fig.~\ref{fig1}b and Fig.~\ref{fig1}c). Absolute intensities for the $\rm{C-D}$ stretching 
modes are 18.128 km/mole, 8.132 km/mole and 11.682 km/mole respectively for the three isomers.

\begin{table*}
\centering
\caption{Theoretical spectral data for Deuteronated pyrene and its isomers}
\label{tab1}
\resizebox{\textwidth}{!}{%
\newcommand{\mc}[3]{\multicolumn{#1}{#2}{#3}}
\begin{tabular}[c]{c|c|c|c|c|ccc|}
\hline
\hline
DPAH$^+$ & Frequency & Wavelength & Relative& \mc{4}{c}{Mode} \\
 & ($\rm cm^{-1}$) & (~$\mu \rm m$) & Intensity & \mc{4}{c}{} \\ \hline
Deuteronated & 700 & 14.29 & 0.05 & \mc{4}{c}{$\rm{C-D}$ in plane + $\rm{C-C-C}$ in plane} \\
Pyrene & 1126 & 8.87 & 0.86 & \mc{4}{c}{$\rm{H-C-D}$ oop + $\rm{C-H}$ in plane} \\
  & 2092 & 4.78 & 0.14 & \mc{4}{c}{$\rm{C-D}$ stretching} \\
 \hline
Deuteronated & 866 & 11.54 & 0.31 & \mc{4}{c}{$\rm{H-C-D}$ in plane + $\rm{C-H}$ oop} \\
Pyrene & 1176 & 8.5 & 0.1 & \mc{4}{c}{$\rm{H-C-D}$ oop + $\rm{C-H}$ in plane} \\
(isomer1) & 1185 & 8.44 & 0.19 & \mc{4}{c}{$\rm{H-C-D}$ oop + $\rm{C-H}$ in plane} \\
\hline
 & 764 & 13.08 & 0.08 & \mc{4}{c}{$\rm{C-D}$ in plane + $\rm{C-C-C}$ in plane + $\rm{C-H}$ oop} \\
Deuteronated & 829 & 12.06 & 0.39 & \mc{4}{c}{$\rm{C-D}$ in plane +$\rm{C-H}$ oop} \\
Pyrene & 858 & 11.65 & 0.11 & \mc{4}{c}{$\rm{C-D}$ in plane + $\rm{C-H}$ oop} \\
(isomer2) & 1163 & 8.6 & 0.4 & \mc{4}{c}{$\rm{H-C-D}$ oop + $\rm{C-H}$ in plane} \\
 & 1171 & 8.54 & 0.1 & \mc{4}{c}{$\rm{H-C-D}$ oop + $\rm{C-H}$ in plane} \\
 & 2103 & 4.76 & 0.07 & \mc{4}{c}{$\rm{C-D}$ stretching} \\
\hline
\end{tabular}
}%
\end{table*}

The $\rm{C-D}$ stretching mode at 4.78~$\mu \rm m$ does not overlap with any other mode and appears as a new feature. 
There is also an aliphatic $\rm{C-H}$ bond in deuteronated pyrene at the addition site of deuteron. A spectral band near 
$\sim~2850~\rm cm^{-1}$ (3.5~$\mu \rm m$) is attributed to aliphatic $\rm{C-H}$ stretching, while those 
near $\sim~3060~\rm cm^{-1}$ (3.3~$\mu \rm m$) are due to aromatic $\rm{C-H}$ stretching. D-associated 
modes for deuteronated pyrene and its isomers are presented in Table \ref{tab1}. Relative intensities 
above 0.05 only are listed.
\begin{figure*}
\centering
\includegraphics[width=10cm,height=15cm]{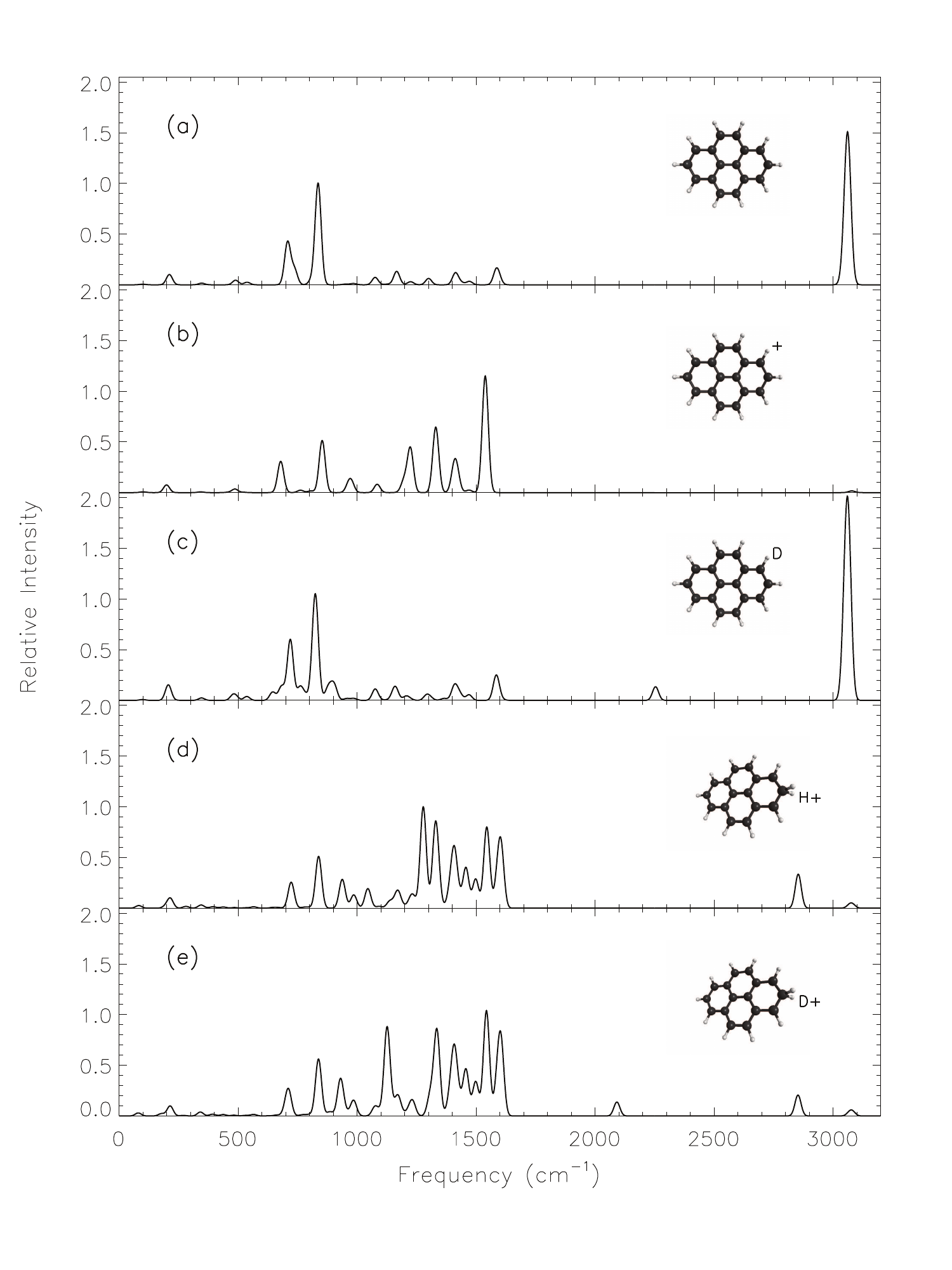}
\caption{ Theoretical spectra of (a) Neutral pyrene, (b) Pyrene cation, (c) Deuterated pyrene,
(d) Protonated pyrene and (e) Deuteronated Pyrene}
\label{fig2}
\end{figure*}

DFT calculations have been carried out on neutral pyrene, cationic pyrene, deuterated pyrene and protonated pyrene 
to compare the spectra with deuteronated pyrene. Fig.~\ref{fig2} shows comparison of these spectra with that of 
deuteronated pyrene. Deuteronated pyrene is transitionally active in the $\sim~900-1600~\rm cm^{-1}$ (11-6~$\mu \rm m$) 
range (Fig.~\ref{fig2}e). This is reasonably similar to protonated pyrene with a variation in intensity (Fig.~\ref{fig2}d). 
For deuteronated pyrene, the intensities of the $\rm{C-H}$ in plane and $\rm{C-C}$ stretching modes are increased by 
a factor of $\sim$~1.2 compared to protonated pyrene. Deuteronated pyrene being structurally similar to protonated 
pyrene shares similar types of vibrational characteristic modes. However, due to the larger mass of deuterium, all
D-associated modes in deuteronated pyrene are red-shifted compared to the protonated form; the $\rm{H-C-D}$ oop mode
shifts from  1278~$\rm cm^{-1}$ (7.82~$\mu \rm m$) to 1126~$\rm cm^{-1}$ (8.87~$\mu \rm 
m$) compared to its protonated counterpart. This particular mode at 8.87~$\mu \rm m$ is distinct 
(Fig.~\ref{fig2}e) and does not appear in any other form of pyrene. Another striking difference is 
the presence of a mode at 2092~$\rm cm^{-1}$ (4.78~$\mu \rm m$) (Fig.~\ref{fig2}e) in deuteronated 
pyrene which is assigned to $\rm{C-D}$ stretching. On comparing with its protonated counterpart, it 
seems that the relative intensity (0.3) of the $\rm{H-C-H}$ symmetric stretching mode at 2854~$\rm cm^{-1}$ 
(3.5~$\mu \rm m$) in protonated pyrene is divided between the $\rm{C-D}$ stretching mode (relative intensity 0.14) 
at 2092~$\rm cm^{-1}$ (4.78~$\mu \rm m$) and the aliphatic $\rm{C-H}$ stretching mode (relative intensity 0.21)
at 2853~$\rm cm^{-1}$ (3.5~$\mu \rm m$) for deuteronated pyrene. Cationic pyrene also shows significant 
transitions in the 900-1600~$\rm cm^{-1}$ (11-6~$\mu \rm m$) region (Fig.~\ref{fig2}b), but there are fewer
modes compared to protonated and deuteronated pyrene. This may be due to reduction in the symmetry for 
deuteronated pyrene. Neutral pyrene and deuterated pyrene show weak features in this region and the $\rm{C-H}$ 
stretching mode at $3050~\rm cm^{-1}$ (3.3~$\mu \rm m$) dominates with high intensity (Fig.~\ref{fig2}a 
and Fig.~\ref{fig2}c). For cationic, protonated and deuteronated pyrene, there are weak bands in the 
3.3~$\mu \rm m$ region. As for deuteronated pyrene, the $\rm{C-D}$ stretching mode also exists for 
deuterated pyrene (Fig.~\ref{fig2}c), but is shifted to shorter wavelength and appears at $2254~\rm cm^{-1}$ (4.44~$\mu \rm m$).

\subsection*{Deuteronated perylene}

\begin{figure*}
\centering
 \includegraphics[width=10cm,height=12cm]{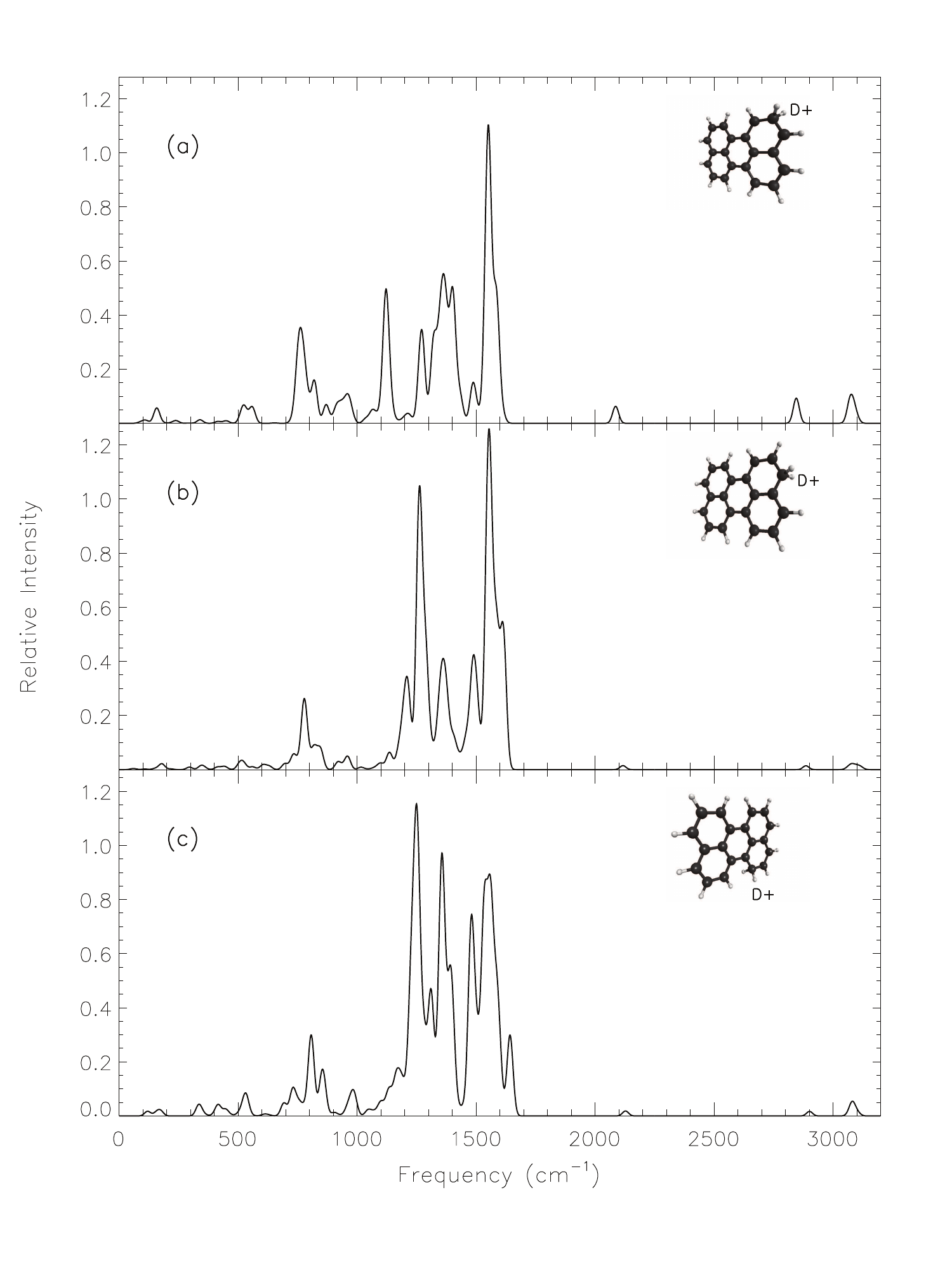}
 \caption{ Theoretical spectra of (a) DC$_{20}\rm H_{12}^+$, (b) Isomer 1 of  DC$_{20}\rm H_{12}^+$,
 (c) Isomer 2 of DC$_{20}\rm H_{12}^+$}
 \label{fig3}
 \end{figure*}

Deuteronated perylene (DC$_{20}\rm H_{12}^+$) has three isomers and the spectra are presented in Fig.~\ref{fig3}. 
All the isomers have C$\rm_s$ symmetry and contain D-contributing features along with the other usual PAH bands.
$\rm{C-D}$ in-plane features are present between $730-870~\rm cm^{-1}$ (13.7-11.5~$\mu \rm m$) and $\rm{H-C-D}$ 
oop in the $1120-1212~\rm cm^{-1}$ (8.9-8.3~$\mu \rm m$) range. The $\rm{C-D}$ stretching mode 
appears at $2087~\rm cm^{-1}$ (4.79~$\mu \rm m$) (Fig.~\ref{fig3}a) and is not distinct in the other 
two isomers (Fig.~\ref{fig3}b \& Fig.~\ref{fig3}c). Absolute intensities for the $\rm{C-D}$ stretching 
mode for the three isomers of deuteronated perylene are 14.073 km/mole, 5.004 km/mole and 4.582 km/mole 
respectively. Bands near $\sim~2850~\rm cm^{-1}$ (3.5~$\mu \rm m$) and $\sim~3060~\rm cm^{-1}$ (3.3~$\mu \rm m$) 
account for aliphatic $\rm{C-H}$ stretching and aromatic $\rm{C-H}$ stretching of deuteronated perylene, 
respectively. Spectral data for D-contributing modes with relative intensities above 0.05 are presented in Table \ref{tab2}.

 \begin{table*}
\centering
\caption{Theoretical spectral data for Deuteronated perylene and its isomers}
\label{tab2}
\resizebox{\textwidth}{!}{%
\newcommand{\mc}[3]{\multicolumn{#1}{#2}{#3}}
\begin{tabular}[c]{c|c|c|c|c|ccc|}
\hline
\hline
DPAH$^+$ & Frequency & Wavelength & Relative& \mc{4}{c}{Mode} \\
  & ($\rm cm^{-1}$) & ($\mu${}m) & Intensity & \mc{4}{c}{} \\ \hline
  & 755 & 13.25 & 0.18 & \mc{4}{c}{$\rm{C-D}$ in plane + $\rm{C-H}$ oop} \\
  & 770 & 12.98 & 0.18 & \mc{4}{c}{$\rm{H-C-D}$ in plane + $\rm{C-H}$ oop} \\
  & 786 & 12.73 & 0.06 & \mc{4}{c}{$\rm{H-C-D}$ in plane + $\rm{C-H}$ oop} \\
  Deuteronated & 870 & 11.49 & 0.07 & \mc{4}{c}{$\rm{H-C-D}$ in plane + $\rm{C-H}$ oop} \\
  Perylene  & 1120 & 8.93 & 0.39 & \mc{4}{c}{$\rm{H-C-D}$ oop + $\rm{C-H}$ in plane} \\
  & 1130  & 8.85 & 0.12 & \mc{4}{c}{$\rm{H-C-D}$ oop + $\rm{C-H}$ in plane} \\
   & 2087 & 4.79 & 0.06 & \mc{4}{c}{$\rm{C-D}$ stretching} \\
   \hline
  & 778 & 12.86 & 0.23 & \mc{4}{c}{$\rm{C-D}$ in plane + $\rm{C-H}$ oop} \\
 Deuteronated & 846 & 11.82 & 0.07 & \mc{4}{c}{$\rm{C-D}$ in plane + $\rm{C-H}$ oop + $\rm{C-C-C}$ in plane} \\
 Perylene & 1197 & 8.35 & 0.1 & \mc{4}{c}{$\rm{H-C-D}$ oop + $\rm{C-H}$ in plane} \\
 (isomer 1) & 1212 & 8.25 & 0.22 & \mc{4}{c}{$\rm{H-C-D}$ oop + $\rm{C-H}$ in plane} \\
 \hline
 Deuteronated & 730 & 13.7 & 0.1 & \mc{4}{c}{$\rm{C-D}$ in plane + $\rm{C-H}$ oop} \\
 Perylene & 853 & 11.73 & 0.13 & \mc{4}{c}{$\rm{C-D}$ in plane + $\rm{C-H}$ oop + $\rm{C-C-C}$ in plane} \\
 (isomer 2) & 862 & 11.59 & 0.05 & \mc{4}{c}{$\rm{C-D}$ in plane + $\rm{C-H}$ oop + $\rm{C-C-C}$ in plane} \\
 \hline
 \end{tabular}
  }%
 \end{table*}

\begin{figure*}
\centering
 \includegraphics[width=10cm,height=15cm]{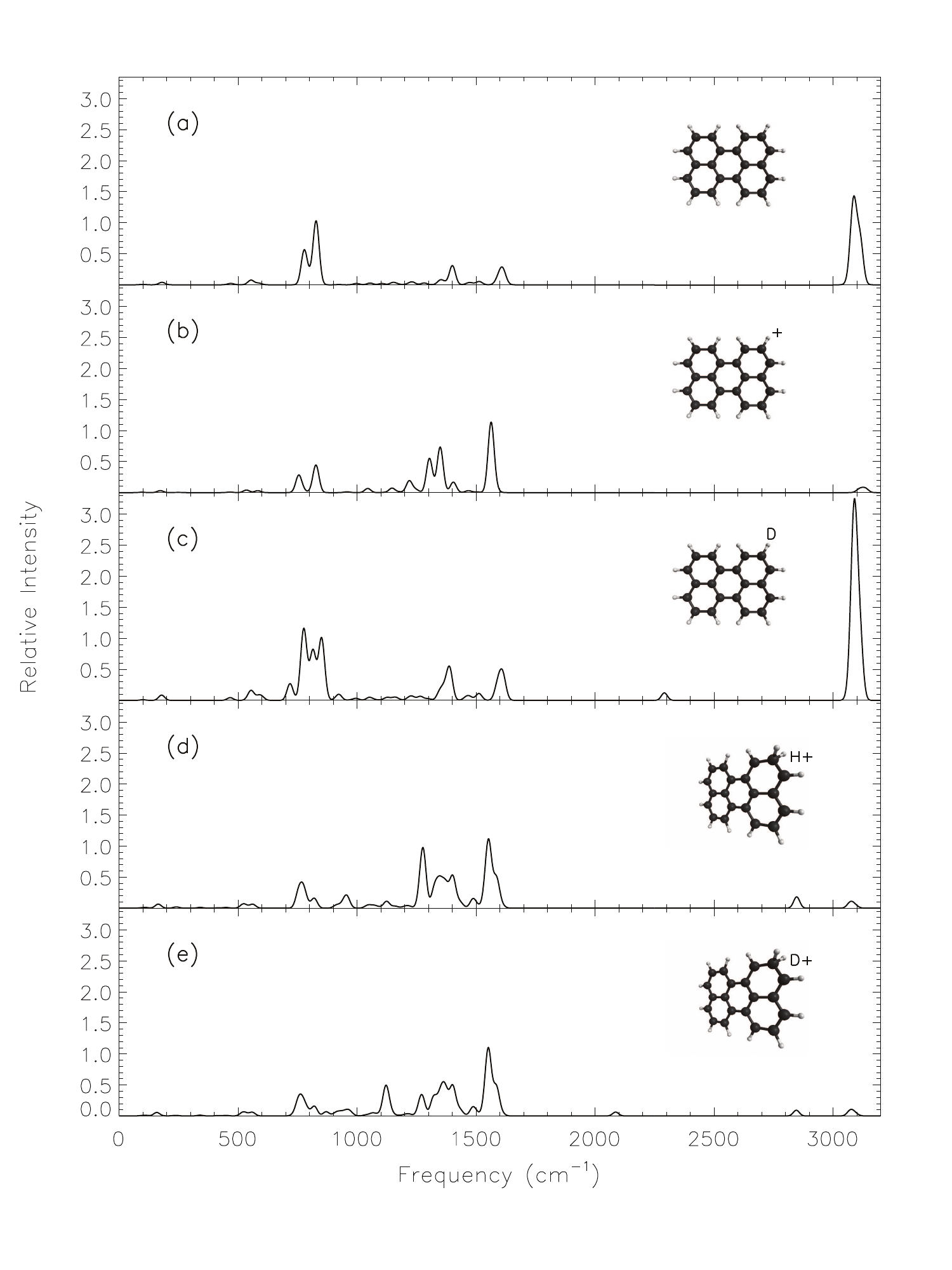}
 \caption{Theoretical spectra of (a) Neutral perylene, (b) Perylene cation, (c) Deuterated perylene,
(d) Protonated perylene and (e) Deuteronated Perylene}
 \label{fig4}
 \end{figure*}

A comparison of the theoretical spectrum of deuteronated perylene with its neutral, cation, deuterated and protonated 
forms is shown in Fig.~\ref{fig4}. Deuteronated and protonated perylene participate in similar types of vibrational modes 
with increasing number of transitions in the $900-1600~\rm cm^{-1}$ (11-6~$\mu \rm m$) range compared to their neutral 
counterparts (perylene and  deuterated perylene). Despite the fact that most of the modes are similar in this region, 
they do vary in intensity. However, variation in intensity is not uniform. Protonated and deuteronated perylene show 
maximum intensity for $\rm{C-C}$ stretch vibration close to $1550~\rm cm^{-1}$ (6.45~$\mu \rm m$) (Fig.~\ref{fig4}d \&  Fig.~\ref{fig4}e). 
For deuteronated perylene, $\rm{H-C-D}$ oop occurs at 1120~$\rm cm^{-1}$ (8.93~$\mu \rm m$) and 1130~$\rm cm^{-1}$ 
(8.85~$\mu \rm m$) mixing with the $\rm{C-H}$ in plane modes (Fig.~\ref{fig4}e). Other forms of perylene do 
not show any significant intensity modes near this wavenumber. Deuterated and deuteronated perylene show
new features at 2291~$\rm cm^{-1}$ (4.36~$\mu \rm m$) and 2087~$\rm cm^{-1}$ 
(4.79~$\mu \rm m$), respectively (Fig.~\ref{fig4}c \&  Fig.~\ref{fig4}e). These two transitions arise due to 
aromatic (deuterated perylene) and aliphatic (deuteronated perylene) $\rm{C-D}$ stretching respectively. All 
forms of perylene show bands near $3050~\rm cm^{-1}$ (3.3~$\mu \rm m$) with varying intensities which is attributed 
to aromatic $\rm{C-H}$ stretching. Neutral forms of perylene (perylene and deuterated perylene) show strong 
intensities at this wavenumber, while cationic, protonated and deuteronated perylene have weak features in 
this region. The presence of an aliphatic $\rm{C-H}$ bond in protonated and deuteronated perylene produces 
features at 2848~$\rm cm^{-1}$ (3.5~$\mu \rm m$) and 2846~$\rm cm^{-1}$ (3.5~$\mu \rm m$) due to stretching 
of the $\rm{C-H}$ bond. It is recognized that the  intensity of aliphatic $\rm{H-C-H}$ stretching mode (0.15) in
protonated perylene is distributed among intensities of aliphatic $\rm{C-D}$ stretching (0.06) and aliphatic 
$\rm{C-H}$ stretching (0.09) in deuteronated perylene.

 \subsection*{Deuteronated coronene}

   \begin{figure*}
 \centering
 \includegraphics[width=10cm, height=15cm]{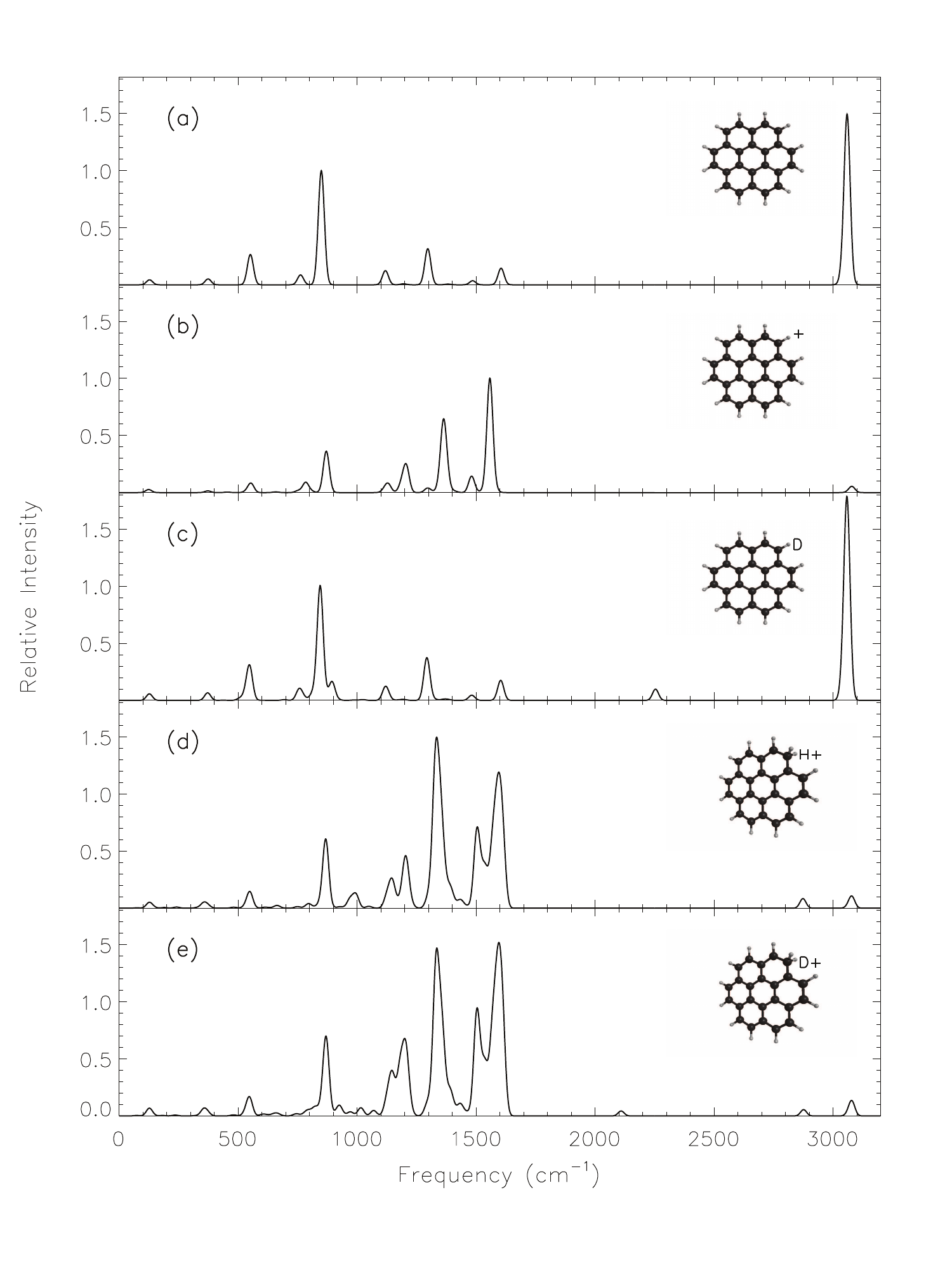}
 \caption{Theoretical spectra of (a) Neutral coronene, (b) Coronene cation, (c) Deuterated coronene,
(d) Protonated coronene and (e) Deuteronated coronene}
 \label{fig5}
 \end{figure*}

Coronene has only one unique site of deuteronation. Fig.~\ref{fig5}e shows the theoretically predicted spectrum for 
deuteronated coronene. The $\rm{C-D}$ in-plane modes are distributed in the $823-871~\rm cm^{-1}$ ($12.15-11.5~\mu \rm m$)
region, the $\rm{H-C-D}$ oop mode is present at $1181~\rm cm^{-1}$ ($8.47~\mu \rm m$) and the $\rm{C-D}$ stretching 
mode appears weakly at $2110~\rm cm^{-1}$ ($4.74~\mu \rm m$). Absolute and relative intensities for the $\rm{C-D}$ stretching mode are 
7.328~km/mole and 0.04 respectively. The spectral data are listed in Table \ref{tab3}.

  \begin{table*}
\centering
\caption{Theoretical spectral data for Deuteronated coronene}
\label{tab3}
\resizebox{.85\textwidth}{!}{%
\newcommand{\mc}[3]{\multicolumn{#1}{#2}{#3}}
\begin{tabular}[c]{c|c|c|c|c|ccc|}
\hline
\hline
  DPAH$^+$ & Frequency & Wavelength & Relative& \mc{4}{c}{Mode} \\
     & ($\rm cm^{-1}$) & ($\mu \rm m$) & Intensity & \mc{4}{c}{} \\ \hline
   & 823 & 12.15 & 0.07 & \mc{4}{c}{$\rm{C-D}$ in plane + $\rm{C-H}$ oop} \\
  Deuteronated & 863 & 11.59 & 0.15 & \mc{4}{c}{$\rm{C-D}$ in plane + $\rm{C-H}$ oop} \\
  Coronene & 871 & 11.48 & 0.6 & \mc{4}{c}{$\rm{C-D}$ in plane + $\rm{C-H}$ oop} \\
   & 1181 & 8.47 & 0.34 & \mc{4}{c}{$\rm{H-C-D}$ oop + $\rm{C-H}$ in plane} \\
   \hline
  \end{tabular}
  }%
  \end{table*}

Fig.~\ref{fig5} also shows the comparison of deuteronated coronene with its neutral, cation, deuterated and protonated 
forms. For deuteronated and protonated coronene, spectral modes in $\sim~1352-1612~\rm cm^{-1}$ ($7.4-6.2~\mu \rm m$) 
region follow a similar pattern with variation in intensity. Intensities for deuteronated coronene are increased by 
an average factor of $\sim~1.4$ compared to its protonated form in this region. The maximum intensity for both 
appears nearly at same position at $1330~\rm cm^{-1}$ ($7.52~\mu \rm m$), but the corresponding modes are different. 
For deuteronated coronene, the peak intensity arises due to the combination of $\rm{C-H}$ in-plane and $\rm{C-C}$ 
stretch modes, whereas, for protonated coronene, it is due to the combination of $\rm{H-C-H}$ oop and $\rm{C-C}$ 
stretch modes. The $\rm{H-C-D}$ oop mode is prominent at $1181~\rm cm^{-1}$ (8.47~$\mu \rm m$) for deuteronated 
coronene (Fig.~\ref{fig5}e). The $\rm{C-D}$ stretching vibrational mode for deuterated and 
deuteronated coronene fall in a featureless region at $2254~\rm cm^{-1}$ (4.44~$\mu \rm m$) and $2110~\rm cm^{-1}$ 
(4.74~$\mu \rm m$) respectively (Fig.~\ref{fig5}c \& Fig.~\ref{fig5}e). These two modes are weak in intensity. 
Cationic coronene (Fig.~\ref{fig5}b) shows a greater number of transitions in the $\rm{C-H}$ in plane and $\rm{C-C}$ stretching 
region (900-1600~$\rm cm^{-1}$) compared to its neutral forms (coronene and deuterated coronene, Fig.~\ref{fig5}a and 
Fig.~\ref{fig5}c respectively). Protonation and deuteronation further increases the number of transitions 
in this region (Fig.~\ref{fig5}d and Fig.~\ref{fig5}e respectively). Cationic, protonated and deuteronated 
forms of coronene show faint features at $3050~\rm cm^{-1}$ ($3.3~\mu \rm m$) unlike the neutral 
counterparts.

From the data presented in Table 1-3, it is deduced that the $\rm{C-D}$ in-plane and $\rm{H-C-D}$ oop modes 
overlap the regions corresponding  to $\rm{C-H}$ oop and $\rm{C-H}$ in-plane modes, respectively. 
The $\rm{C-D}$ stretching modes do not overlap with any other mode and  are identified easily. For 
all three molecules along with their isomers (discussed above), $\rm{C-D}$ in plane modes are found
to be distributed in the range $\sim~700-870~\rm cm^{-1}$ ($14-11~\mu \rm m$) with a range of 
intensities. The $\rm{H-C-D}$ oop modes appear in the narrow range $\sim~1120-1212~\rm cm^{-1}$ 
($8.9-8.3~\mu \rm m$). A less intense feature is seen at $\sim~2105~\rm cm^{-1}$ (4.75~$\mu \rm m$) 
which arises due to the $\rm{C-D}$ stretching mode of DPAH$^+$. Symmetrical deuteronated pyrene 
(C$\rm_{2v}$) shows a greater number of transitions compared to deuteronated perylene (C$\rm_s$) 
and deuteronated coronene (C$\rm_s$).

\citet[]{Peeters04} reported the detection of PAD features at $4.4~\mu \rm m$ and $4.65~\mu \rm m$ towards the 
Orion Nebula and M17 with the use of \textit{SWS} on board \textit{ISO}. \citet[]{Onaka14} using \textit{AKARI} 
observed an overlapping region but did not confirm the detection and suggested the presence of similar bands 
with much weaker intensity. Moreover, \citet[]{Onaka14} did not look at the excess around 4.75~$\mu \rm m$ region.
The observations of \citet[]{Peeters04} covered a larger area of Orion and M17 compared to the observations 
of \citet[]{Onaka14} owing to the large aperture of the \textit{SWS} ($20'' \times 14''$) on-board the \textit{ISO} compared to the 
\textit{IRC} (Ns~$ - 0'.8 \times 5''$ and Nh~$ - 1' \times 3''$) on \textit{AKARI}. The detections of the 4.4 and 4.65~$\mu \rm m$ bands 
reported by \citet[]{Peeters04} have large uncertainties and are at a level of 4.4 $\sigma$ and 1.9 $\sigma$ respectively in M17 and Orion. Stretching of the $\rm{C-D}$ aromatic and $\rm{C-D}$ aliphatic 
bonds in PADs produce bands close to $4.4~\mu \rm m$ and $4.65~\mu \rm m$ respectively. These bands are analogous 
to the 3.3 and $3.5~\mu \rm m$ bands which arise due to aromatic and aliphatic $\rm{C-H}$ stretching respectively. From their observations, 
\citet[]{Peeters04} calculated the D/H ratio ($0.17\pm0.03$ towards the Orion bar and $0.36\pm0.08$ towards M17) 
by taking the ratio of the sum of intensities for bands in the $4-5~\mu \rm m$ ($\rm{C-D}$ stretching) region 
to the analogous sum near $3.3~\mu \rm m$ band ($\rm{C-H}$ stretching). \citet[]{Onaka14} proposed 
a significantly smaller D/H ratio of $0.02-0.03$.

 \section{Theoretical spectra for deuterated-deuteronated coronene and deuteronated circumcoronene}

 \begin{figure*}
 \centering
 \includegraphics[width=10cm, height=6cm]{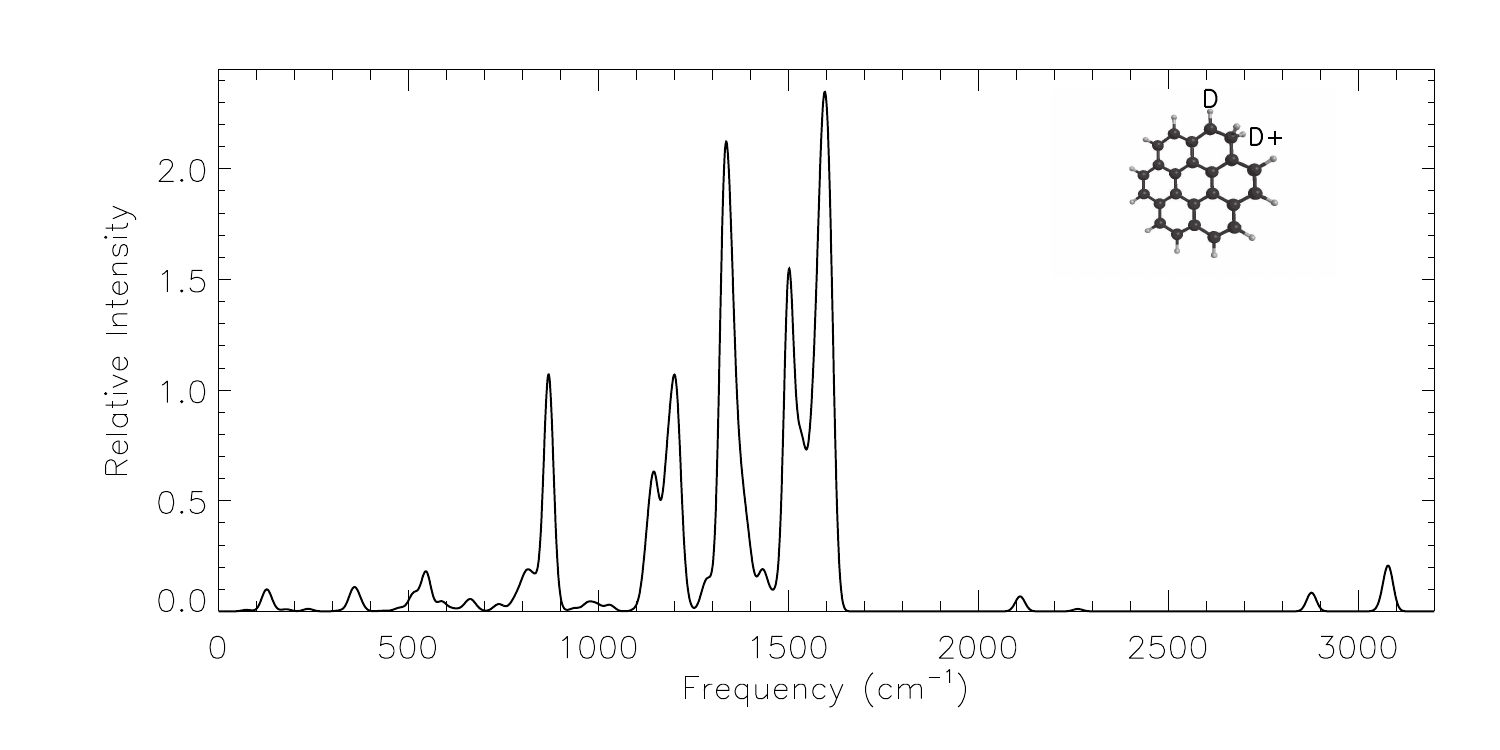}
 \caption{Theoretical spectrum of deuterated-deuteronated coronene (DcorD$^+$)}
 \label{fig6}
 \end{figure*}

 Fig.~\ref{fig6} shows the calculated spectrum of deuterated-deuteronated coronene (DcorD$^+$). In DcorD$^+$, a deuterium 
 atom replaces  a hydrogen atom to form a $\rm{C-D}$ bond and a deuteron (D$^+$) is added to a $\rm{C-H}$ site of the 
 neutral coronene (C$_{24}$H$_{12}$). The resulting DcorD$^+$ molecule thus carries two types of $\rm{C-D}$ bond, 
 `aromatic' at the addition site of D and aliphatic at the addition site of D$^+$. The spectrum is very similar to
 that of deuteronated coronene except for a variation in intensity in the $\sim1318-1570~\rm cm^{-1}$ ($7.6-6.4~\mu \rm m$) 
 region. There are weak bands at $2110~\rm cm^{-1}$ ($4.7~\mu \rm m$), $2261~\rm cm^{-1}$ ($4.4~\mu \rm m$) and 
 $2876~\rm cm^{-1}$ ($3.5~\mu \rm m$) with relative intensities of 0.07, 0.01 and 0.08 respectively. Absolute 
 intensities for these particular bands are 7.4811 km/mole, 1.2101 km/mole and 9.3345 km/mole respectively. The 
 $4.4~\mu \rm m$ feature is barely visible in the spectrum. As discussed earlier, these transitions are 
 characteristic of aliphatic $\rm{C-D}$ stretching, aromatic $\rm{C-D}$ stretching and aliphatic $\rm{C-H}$ 
 stretching, respectively.

\begin{figure*}
 \centering
 \includegraphics[width=10cm, height=6cm]{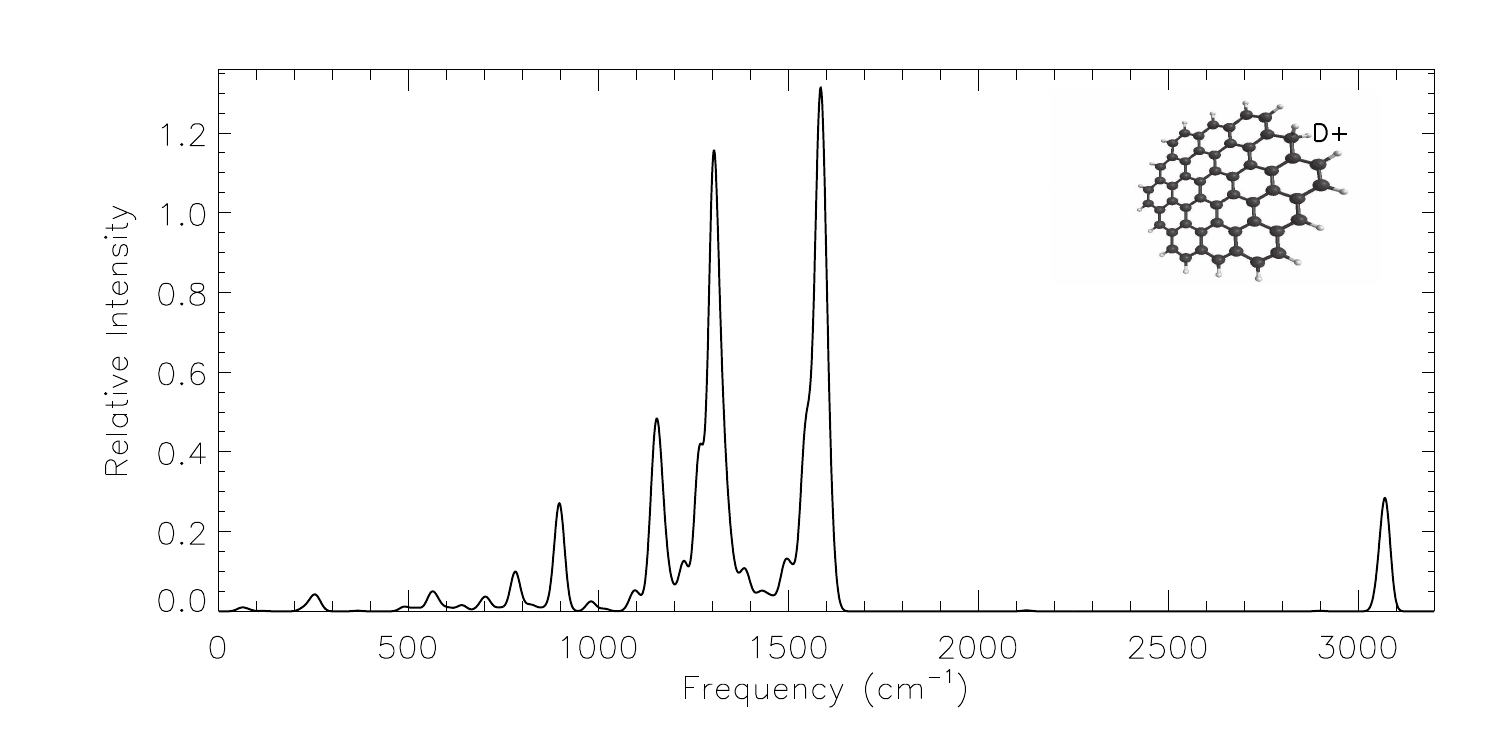}
 \caption{Theoretical spectrum of deuteronated circumcoronene}
 \label{fig7}
 \end{figure*}

DFT calculations were also performed on a large-sized DPAH$^+$, i.e. deuteronated circumcoronene (DC$_{54}$H$_{18}^+$), 
to investigate the effect of PAH size on the intensity and position of the $\rm{C-D}$ stretching mode. As shown in 
Fig.~\ref{fig7}, the $\rm{C-D}$ stretching mode at $\sim~2127~\rm cm^{-1}$ ($4.7~\mu \rm m$) for deuteronated 
circumcoronene is too weak to be visible. The absolute and relative intensity for this mode are 1.3165~km/mole 
and 0.003, respectively.

With increase in size of DPAH$^+$, the absolute and relative intensities of the $\rm{C-D}$ stretching transitions and 
the [D/H]$\rm_{num}$ ratio (number of D atoms/number of H atoms) decrease (Table~\ref{tab4}). DcorD$^+$ has two D 
atoms unlike other DPAH$^+$s considered and hence has a higher [D/H]$\rm_{num}$ ratio of 0.18. Another point to note 
is that as the size of the DPAH$^+$ increases, the band corresponding to the aliphatic $\rm{C-D}$ stretching mode 
shifts towards shorter wavelengths and closer to the observed astronomical band position of $4.65~\mu \rm m$. The 
absolute intensity, relative intensity and the position of the aliphatic $\rm{C-D}$ stretching mode for deuteronated
pyrene, deuteronated perylene, deuteronated coronene, DcorD$^+$ and deuteronated circumcoronene are compared in 
Table \ref{tab4}. [D/H]$\rm_{int}$ ratio has been determined by taking the ratio of the intensities of the $\rm{C-D}$ 
stretch bands to the intensities of the $\rm{C-H}$ stretch bands as obtained from the calculations. 
There is no correlation present between [D/H]$\rm_{int}$ and [D/H]$\rm_{num}$. Having the same [D/H]$\rm_{num}$ 
ratio, deuteronated perylene and deuteronated coronene give [D/H]$\rm_{int}$ ratios of nearly same magnitude. 
DcorD$^+$ having two D atoms shows a higher [D/H]$\rm_{int}$ ratio of 0.26 compared to that of deuteronated 
coronene (one D atom). An exception is deuteronated circumcoronene for which the [D/H]$\rm_{int}$ ratio is 
lower by an order of magnitude compared to [D/H]$\rm_{num}$ ratio. [D/H]$\rm_{int}$ ratio is dependent on the
size of PAHs and percentage of deuteronation. Hence, for a comparative analysis, we compute [D/H]$\rm_{sc}$ 
which is [D/H]$\rm_{int}$ per [D/H]$\rm_{num}$ (Table~\ref{tab4}). [D/H]$\rm_{sc}$ is found to be decreasing
with increase in size of the molecule.

  \begin{table*}
 \centering
  \begin{minipage}{140mm}
 \caption{Intensities and positions of aliphatic $\rm{C-D}$ stretching mode in DPAH$^+$s}
 \label{tab4}

 \begin{tabular}[c]{c|c|c|c|c|c|c|c}
 \hline
 \hline
 DPAH$^+$ & Frequency & Wavelength & Absolute & Relative & [D/H]$\rm_{num}$ \footnote{[D/H]$\rm_{num}=$no of D atoms/no of H atoms} %
 & [D/H]$\rm_{int}$ \footnote{[D/H]$\rm_{int}=$intensity of $\rm{C-D}$ stretch/intensity of $\rm{C-H}$ stretch} 
 & [D/H]$\rm_{sc}$ \footnote{[D/H]$\rm_{sc}$=$\frac{\rm [D/H]_{int}}{\rm [D/H]_{num}}$}  \\
  & ($\rm cm^{-1}$) & ($\mu \rm m$) & Intensity (km/mole) & Intensity &  & \\ \hline
 Deuteronated & 2092 & 4.78 & 18.1282 & 0.138 & 0.10 & 0.500 & 5.00 \\
 Pyrene &  &  &  &  &  &\\
 \hline
 Deuteronated & 2087 & 4.79 & 14.0726 & 0.063 & 0.08 & 0.271 & 3.39\\
 Perylene &  &  &  &  \\ \hline
 Deuteronated & 2110 & 4.74 & 7.3279 & 0.044 & 0.08 & 0.217 & 2.71\\
 Coronene &  &  &  & &  & \\
 \hline
 DcorD$^+$ & 2110 & 4.74 & 7.4811 & 0.068 & 0.18 & 0.258 & 1.43\\ \hline
 Deuteronated &  2127 & 4.70 & 1.3165 & 0.003  & 0.06 & 0.008 & 0.13 \\
 Circumcoronene &  &  &  & &  &\\ \hline
 \end{tabular}
 \end{minipage}
  \end{table*}

 \section{DPAH$^+$ molecules as carriers of UIR emission features}

From the discussion above, it is established that deuterium containing PAHs (PADs and DPAH$^+$s) show features in the 
$4-5~\mu \rm m$ region. This is the same region where spectral contribution of deuterated PAHs have been discussed by \citet[]{Peeters04} and more recently by \citet[]{Onaka14}. This region is featureless for PAHs without deuterium. 
The position of the $\rm{C-D}$ stretch (aromatic and aliphatic) in PADs
and DPAH$^+$s are close to the $4.4$ and $4.65~\mu \rm m$ emission bands observed towards the Orion Nebula and 
M17 \citep[]{Peeters04}. The $4.65~\mu \rm m$ band in DPAH$^+$ is accompanied by a transition at $3.5~\mu \rm m$ 
corresponding to the aliphatic $\rm{C-H}$ stretch. Thus, a condition for DPAH$^+$s to be present in the ISM is that 
the bands at $4.65~\mu \rm m$ and $3.5~\mu \rm m$ should be observed together in the emission spectra of an 
astronomical source. In the emission spectra of Orion nebula and M17 \citep[]{Peeters04}, these two bands have 
indeed been detected which is an indicative of the presence of DPAH$^+$ molecules. In the M17 spectra, \citet[]{Peeters04} have reported the detection of only the $4.65~\mu \rm m$ band at 4.4 $\sigma$ level (the detection of the $4.4~\mu \rm m$ feature towards Orion is with much higher uncertainty). This is a tentative yet strong evidence in support of the presence of deuteronated-PAHs and / or aliphatic deuterated-PAHs. The presence of an aliphatic $\rm{C-D}$ bond results in a band at $4.65~\mu \rm m$ rather than the $4.4~\mu \rm m$ feature which arises due to an aromatic $\rm{C-D}$ stretch vibration. The aliphatic bond also results in features near $3.5~\mu \rm m$ that may indicate the presence of aliphatic $\rm{C-H}$ bonds in neutral and ionized (protonated) PAHs.

In this work we have focused on comparing the band position of DPAH$^+$ molecules with observations rather than correlating 
the intensity of the bands. Therefore, we have not taken into account the effect of excitation of the bands and their intensity.
However, for a direct comparison with the observed spectra, the excitation mechanisms have to be considered. The $\rm{C-D}$ stretch bands being at lower wavenumbers are easily excited compared to the $\rm{C-H}$ stretch bands, thus, care has to be taken while comparing the theoretical and the observed D/H ratios. \citet[]{Onaka14} calculated the emission intensity considering the effect of excitation based on a PAH emission model by \citet[]{Mori12} and found that the smaller cross-section of the $\rm{C-D}$ stretch bands is compensated by its easier excitation compared to $\rm{C-H}$ stretch vibrations. They reported that excitation does not affect the result significantly but an overestimation of D/H ratio by tens of percent is present. We have computed [D/H]$\rm_{sc}$ for DPAH$^+$ molecules, which is nothing but the ratio of $\rm C-D$ stretch and $\rm C-H$ stretch bands per [D/H]$\rm_{num}$.

The observational [D/H] values are estimated by assuming that the band strength per bond is constant for 
the $\rm{C-H}$ and $\rm{C-D}$ bonds. This assumption may not hold for the $\rm{C-D}$ bond. These values are 
compared to the theoretically calculated [D/H] values of specific molecules. The [D/H]$\rm_{sc}$ 
values for deuteronated pyrene, deuteronated perylene, deuteronated coronene and DcorD$^+$ do not fall 
in the range of the D/H value given by observations \citep[]{Peeters04, Onaka14}. With 
increase in the size of DPAH$^+$s, the [D/H]$\rm_{sc}$ value tends to decrease. Observations reported 
by \citet[]{Peeters04} suggest D/H values of 0.17$\pm$0.03 in the Orion bar and 0.36$\pm$0.08 in M17.
However, \citet[]{Onaka14} estimate a significantly smaller D/H value of $0.03$ which is an order of
magnitude smaller than the value proposed by \citet[]{Peeters04}; which points 
to the fact that if Ds are depleted onto PAHs, they might be accommodated in large PAHs \citep[]{Onaka14}.
Large PAHs have a tendency to be ionized in the ISM which may subsequently add a deuterium to form DPAH$^+$.
Therefore, in such interstellar regions, the formation of DPAH$^+$s may be preferred over PADs. The D/H
ratios proposed by \citet[]{Peeters04} and \citet[]{Onaka14} may be used to estimate the size of PADs or 
DPAH$^+$s in the ISM. From this work, deuteronated circumcoronene shows a [D/H]$\rm_{sc}$ value close 
to \citet[]{Peeters04} observation. For the smaller deuteronated PAHs, the [D/H]$\rm_{sc}$ values are
higher by about an order of magnitude. It is noted that DPAH$^+$ molecules of the size of circumcoronene 
(54 carbon atoms) and larger satisfy the D to H ratio as observed by \citet[]{Peeters04} and \citet[]{Onaka14}. 
This points to the fact the DPAH$^+$ molecules if present may have 50 or more 
carbon atoms.

Following the observed band ratio as found by \citet[]{Onaka14}, the DPAH$^+$ molecules present in the ISM should be large. 
If this assumption is strictly followed, only large DPAH$^+$ molecules with low values of [D/H]$\rm_{num}$ will exist in the ISM and these will not be able to account for the inferred depletion of D \citep[]{Draine06}. 
The expected D/H without depletion is around 20 parts per million (ppm) and the minimum observed ratio is about 7 ppm. 
If this difference is attributed solely to the depletion of D onto PAHs then
the necessary concentration of D/H (value of [D/H]$\rm_{num}$) in PAHs would be 0.3. This points to the fact that large PAHs that match the D/H value estimated from observed band ratios will have lower elemental D/H values and therefore, will not match the depletion model values of \citet[]{Draine06}. This discrepancy warrants for an extensive observational programme for the search of DPAH$^+$ and PAD molecules in the ISM. Theoretical and experimental spectroscopic studies are further needed to complement the observations. A refined depletion model of D on PAHs may thus be obtained based on the results of observations and spectroscopic studies.

\section{Conclusion}

Interstellar PAHs may have significant deuterium content with a D/H ratio as high as 0.3 \citep[]{Draine06}. In 
this context, we have calculated the vibrational spectra of deuteronated PAHs and have compared them with those of the corresponding neutral, cationic, deuterated and protonated forms. The theoretical spectral 
data provides strong evidences in support of DPAH$^+$ molecules to be part of the interstellar PAH family and 
these may be responsible for some of the observed IR features. In particular, the generally featureless 
region of $4-5~\mu \rm m$ of pure PAHs may be dominated by features due to $\rm{C-D}$ bond vibrations in 
DPAH$^+$s or deuterated PAHs. Stretching of the aliphatic $\rm{C-D}$ bond gives a distinct transition near 
$4.7~\mu \rm m$. This warrants for a look at the excess around $4.7~\mu \rm m$ region in astronomical sources.

Our calculations of deuteronated circumcoronene yield a D/H ([D/H]$\rm_{sc}$) value that is similar to ones obtained by \citet[]{Peeters04}.
Large DPAH$^+$s will have even small D/H values that might be in accordance with the proposed value of \citet[]{Onaka14}.
A higher deuterium fraction has been observed in M17 \citep[]{Peeters04} with lower uncertainty which only shows the
$4.65~\mu \rm m$ band (the aliphatic deuterium bond) and not the $4.4~\mu \rm m$ band (the aromatic deuterium bond).
This clearly points to the dominance of deuteronated-PAHs compared to deuterated-PAHs in such regions.

This study may also be used as input to deuterium depletion models and also for estimating the  HD/H$_2$ ratio in 
interstellar gas. However for a more conclusive analysis, extensive observations followed by laboratory experiments
are desired.  Revisiting some of the \textit{ISO}, \textit{AKARI} and \textit{SPITZER} data focusing on PAD and DPAH$^+$ systems may provide further insights. Upcoming \textit{James Webb Space Telescope} may supplement this study by the addition of high quality data.

\section*{Acknowledgements}
MB is a junior research fellow in a SERB – DST FAST TRACK project. AP acknowledges financial support from SERB – DST 
FAST TRACK grant (SERB/F/5143/2013-2014) and AP, TO and IS acknowledge support from the DST – JSPS grant (DST/INT/JSPS/P-189/2014). AP thanks the Inter-University Centre for Astronomy and Astrophysics, Pune for associateship. PJS thanks the Leverhulme Trust for award of a Research Fellowship and Leiden Observatory for hospitality that allowed completion of this work.

\def\aj{AJ}%
\def\actaa{Acta Astron.}%
\def\araa{ARA\&A}%
\def\apj{ApJ}%
\def\apjl{ApJ}%
\def\apjs{ApJS}%
\def\ao{Appl.~Opt.}%
\def\apss{Ap\&SS}%
\def\aap{A\&A}%
\def\aapr{A\&A~Rev.}%
\def\aaps{A\&AS}%
\def\azh{AZh}%
\def\baas{BAAS}%
\def\bac{Bull. astr. Inst. Czechosl.}%
\def\caa{Chinese Astron. Astrophys.}%
\def\cjaa{Chinese J. Astron. Astrophys.}%
\def\icarus{Icarus}%
\def\jcap{J. Cosmology Astropart. Phys.}%
\def\jrasc{JRASC}%
\def\mnras{MNRAS}%
\def\memras{MmRAS}%
\def\na{New A}%
\def\nar{New A Rev.}%
\def\pasa{PASA}%
\def\pra{Phys.~Rev.~A}%
\def\prb{Phys.~Rev.~B}%
\def\prc{Phys.~Rev.~C}%
\def\prd{Phys.~Rev.~D}%
\def\pre{Phys.~Rev.~E}%
\def\prl{Phys.~Rev.~Lett.}%
\def\pasp{PASP}%
\def\pasj{PASJ}%
\def\qjras{QJRAS}%
\def\rmxaa{Rev. Mexicana Astron. Astrofis.}%
\def\skytel{S\&T}%
\def\solphys{Sol.~Phys.}%
\def\sovast{Soviet~Ast.}%
\def\ssr{Space~Sci.~Rev.}%
\def\zap{ZAp}%
\def\nat{Nature}%
\def\iaucirc{IAU~Circ.}%
\def\aplett{Astrophys.~Lett.}%
\def\apspr{Astrophys.~Space~Phys.~Res.}%
\def\bain{Bull.~Astron.~Inst.~Netherlands}%
\def\fcp{Fund.~Cosmic~Phys.}%
\def\gca{Geochim.~Cosmochim.~Acta}%
\def\grl{Geophys.~Res.~Lett.}%
\def\jcp{J.~Chem.~Phys.}%
\def\jgr{J.~Geophys.~Res.}%
\def\jqsrt{J.~Quant.~Spec.~Radiat.~Transf.}%
\def\memsai{Mem.~Soc.~Astron.~Italiana}%
\def\nphysa{Nucl.~Phys.~A}%
\def\physrep{Phys.~Rep.}%
\def\physscr{Phys.~Scr}%
\def\planss{Planet.~Space~Sci.}%
\def\procspie{Proc.~SPIE}%
\let\astap=\aap
\let\apjlett=\apjl
\let\apjsupp=\apjs
\let\applopt=\ao

\bibliographystyle{mn} 
\bibliography{mridu}

\end{document}